\documentclass[floats,floatfix,showpacs,amssymb,prd,twocolumn,superscriptaddress,nofootinbib,nolongbibliography,reprint]{revtex4-2}

\usepackage{amssymb,amsmath,verbatim,mathtools,needspace,enumitem,etoolbox,graphicx,physics,microtype,afterpage,xspace,tabularx,lmodern,multirow}
\usepackage{gensymb}
\usepackage[dvipsnames, usenames]{xcolor}
\definecolor{linkcolor}{rgb}{0.0,0.3,0.5}
\usepackage[unicode, colorlinks=true, linkcolor=linkcolor, citecolor=linkcolor, filecolor=linkcolor, urlcolor=linkcolor, linktocpage, breaklinks]{hyperref}
\usepackage[all]{hypcap}
\usepackage[T1]{fontenc}
\usepackage[utf8]{inputenc}
\usepackage[usenames,dvipsnames]{xcolor}
\hypersetup{colorlinks=true,citecolor=romared,linkcolor=romared,urlcolor=romared}

\definecolor{romared}{RGB}{142,0,28}

\newcommand{\be}{\begin{equation}}
\newcommand{\ee}{\end{equation}}

\def\be{\begin{equation}}
\def\ee{\end{equation}}
\newcommand{\beq}{\begin{eqnarray}}
\newcommand{\eeq}{\end{eqnarray}}

\newcommand{\lp}{\left (}
\newcommand{\rp}{\right )}

\newcommand{\PBH}{\text{\tiny PBH}}

\usepackage{aas_macros}

\renewcommand{\d}{{\rm d}}

\newcolumntype{Y}{>{\centering\arraybackslash}X}

\newcommand{\sapienza}{Dipartimento di Fisica, Sapienza Università 
	di Roma, Piazzale Aldo Moro 5, 00185, Roma, Italy}
\newcommand{\infn}{INFN, Sezione di Roma, Piazzale Aldo Moro 2, 00185, Roma, Italy}
\newcommand{\jhu}{\affiliation{Department of Physics and Astronomy, Johns Hopkins University, 3400 North Charles
Street, Baltimore, Maryland 21218, USA}}

\begin{document}
\title{
Primordial black hole mergers from three-body interactions
}

\begin{abstract}
Current gravitational-wave observations set the most stringent bounds on the abundance of primordial black holes (PBHs) in the solar mass range. 
This constraint, however, inherently relies on the merger rate predicted by PBH models. 
Previous analyses have focused mainly on two binary formation mechanisms: 
early Universe assembly out of decoupling from the Hubble expansion and dynamical capture in present-day dark matter structures.
Using reaction rates of three-body processes studied in the astrophysical context,  we show that, under conservative assumptions, three-body interactions in PBH halos efficiently produce binaries. 
Those binaries form at high redshift in Poisson-induced PBH small-scale structures and a fraction  is predicted to coalesce and merge within the current age of the Universe, at odds with the dynamical capture scenario where they merge promptly.
In general, we find that this channel predicts rates comparable to the dynamical capture scenario. 
However, binaries formed from three-body interactions 
do not significantly contribute to the overall PBH merger rate unless
PBHs made up a dominant fraction of the dark matter above the solar mass range, a scenario that is ruled out by current constraints.
Our results support strong bounds on the PBH abundance in the stellar mass range derived from Laser Interferometer Gravitational-Wave Observatory/Virgo/KAGRA observations. 
Finally, we show that both dynamical channels are always subdominant compared to early Universe assembly for PBH mergers in the asteroid mass range, while we expect it to become relevant in scenarios where PBHs are initially strongly clustered.
\end{abstract}

\author{Gabriele Franciolini}
\email{gabriele.franciolini@uniroma1.it}
\affiliation{\sapienza}
\affiliation{\infn}

\author{Konstantinos Kritos}
\email{kkritos1@jhu.edu}
\jhu

\author{Emanuele Berti}
\email{berti@jhu.edu}
\jhu

\author{Joseph Silk}
 	\email{silk@iap.fr}
\jhu
\affiliation{Institut d’Astrophysique de Paris, UMR 7095 CNRS and UPMC, Sorbonne Universit$\acute{e}$, F-75014 Paris, France}
\affiliation{Department of Physics, Beecroft Institute for Particle Astrophysics and Cosmology, University of Oxford, Oxford OX1 3RH, United Kingdom}

\date{\today}
\maketitle

\section{Introduction}\label{sec:intro}

Primordial black holes (PBHs) forming from the collapse of large density fluctuations right after the big bang have been hypothesized to account for a fraction of the dark matter (DM) \cite{Zeldovich:1967lct,Hawking:1974rv,Chapline:1975ojl,Carr:1975qj}. 
Interest in PBHs was reinforced following the first detection of gravitational waves
(GWs) 
by the Laser Interferometer Gravitational-Wave Observatory (LIGO) originating from the merger of two compact objects of around 30 solar masses \cite{LIGOScientific:2016aoc}. 
Calculation of the merger rate from binary PBHs (BPBHs) and comparison with the value inferred from experiment has been performed in the context of early- and late-time Universe BPBHs
\cite{Bird:2016dcv,Clesse:2016vqa,Sasaki:2016jop,Eroshenko:2016hmn, Wang:2016ana, Ali-Haimoud:2017rtz,Chen:2018czv,Raidal:2018bbj}
and constraints on the abundance of PBHs have been set to respect the observed rates of GW detections
\cite{Vaskonen:2019jpv, Gow:2019pok,Wu:2020drm,DeLuca:2020qqa,
Bhagwat:2020bzh,Hall:2020daa,DeLuca:2020jug,Wong:2020yig,Hutsi:2020sol,Kritos:2020wcl,Deng:2021ezy,Kimura:2021sqz,DeLuca:2021wjr,Bavera:2021wmw,Chen:2021nxo,Franciolini:2021tla,Mukherjee:2021ags,Bagui:2021dqi} (see~\cite{Sasaki:2018dmp,Green:2020jor,Franciolini:2021nvv} for reviews).
Interestingly, the LIGO/Virgo/KAGRA Collaboration (LVKC) still allows for a PBH subpopulation with mass scale at around 30$M_\odot$ to contribute to the detections~\cite{Franciolini:2021tla}, 
while being a compelling explanation for the detected mass-gap events \cite{Clesse:2020ghq,DeLuca:2020sae} (such as GW190814~\cite{LIGOScientific:2020zkf} appearing in the hypothesized low mass gap between neutron stars and BHs, and GW190521~\cite{LIGOScientific:2020iuh,LIGOScientific:2020ufj} above the pair instability
limit), which otherwise are  challenging to explain in the standard astrophysical scenarios. Future GW experiments may be able to test such scenarios and constrain the PBH abundance compared to the DM in the solar mass range below ${\cal O}(10^{-5})$ \cite{Ng:2022agi}
(see also Refs.~\cite{DeLuca:2021hde,Franciolini:2021xbq,Pujolas:2021yaw,Mukherjee:2021itf,LISACosmologyWorkingGroup:2022jok,Martinelli:2022elq}).

In most of the literature mentioned above, dynamical capture (denoted by subscript ``cap'' in the following) has been the favored BPBH assembly channel for binaries that form in the late-time Universe. 
This scenario assumes the close encounter of two individual PBHs that interact in a very small region and the BPBH arises via strong emission of GWs. This is the relativistic analog of a two-star tidal capture. 
However, it is well known in the astrophysical literature that other binary assembly scenarios dominate the formation of stellar binaries under certain conditions (see Refs.~\cite{Mandel:2018hfr,Mapelli:2021taw} for recent reviews). 
In particular, binary black hole formation through dynamical channels is expected to be much more efficient in the dense environments of star clusters as opposed to the lower density fields at galactic scales. 
Competing binary formation channels are three-body (hereafter ``3b'') interactions, with one of the members 
removing the amount of energy 
necessary to induce a bound system between the other two bodies. 
In this case, we would have the formation of a BPBH from the triple encounter of three single PBHs.

The velocity dispersion is a crucial parameter in 3b interactions, because whether the third body can remove enough energy to induce the formation of a binary strongly depends on whether the bodies participating in the interaction are moving fast enough for gravitational focusing to dominate the interaction cross section.
Being characterized by a smaller velocity dispersion, strong interactions among objects in smaller systems become more frequent and the probability for binary formation is enhanced. 
To account for the formation of BPBHs in small-scale structures, a crucial ingredient to consider is the
inevitably 
enhanced hierarchical formation of minihalos (with a number of PBH members ranging from a few up to millions) inherited by the Poisson fluctuations in the PBH density field~\cite{Inman:2019wvr,DeLuca:2020jug,Kadota:2020ahr}. 
One may describe such a small-scale structure adopting the Press-Schechter formalism \cite{1974ApJ...187..425P} as customarily done in cosmology to describe the properties of matter distribution in the $\Lambda$ cold dark matter scenario, see e.g. Ref.~\cite{Zentner:2006vw} for a review.
This modeling was confirmed by the  cosmological $N$-body simulations of Ref.~\cite{Inman:2019wvr} and recently used to estimate the effect of PBH clustering properties on constraints from  microlensing searches~\cite{Petac:2022rio,Gorton:2022fyb}
and first star formation~\cite{Liu:2022okz}.
PBH clusters collapse and decouple from the Hubble expansion starting from the onset of the matter-dominated era (i.e., around redshift  $z\lesssim z_\text{\tiny eq}$), and they form dense, virialized halos where BPBH dynamical assembly can occur.

Reference~\cite{Kritos:2020wcl} already hints that the 3b formation channel could dominate over two-body captures in dense astrophysical environments. 
Moreover, according to $N$-body simulations performed in Ref.~\cite{Korol:2019jud}, 3b interactions supply the most dominant BPBH dynamical formation channel in low-mass clusters composed of PBHs.
Here, we estimate the present-day cosmological merger rate density of BPBHs assembled dynamically through 3b interaction.
We will first consider dynamical assembly
in PBH minihalos that start forming in the matter-dominated phase after the epoch of recombination under the assumption that PBHs account for a large fraction of the DM in the Universe. 
For concreteness, we will first consider a PBH population with a narrow mass distribution of PBHs centered around $m_\text{\tiny PBH}=30M_\odot$, 
and then describe how such merger rate scales by varying assumptions on the PBH mass and abundance. It was argued that PBHs formed in the early Universe in the standard scenario have negligible spin \cite{Mirbabayi:2019uph,DeLuca:2019buf,Chongchitnan:2021ehn}. 
Therefore, to further simplify the analysis, we consider nonspinning PBHs.

The paper is organized as follows. In Sec.~\ref{sec:binformc}, we introduce the dynamical formation channel through two-body capture and 3b interactions.
In Sec.~\ref{sec:fraction}, we derive the fraction of binaries which are able to merge within the age of the Universe, given the initial distribution of semimajor axis and eccentricity predicted by the 3b channel. In Sec.~\ref{sec:mini-halos}, we evaluate the contribution from the 3b channel 
in the PBH small-scale structure and compare it to the present epoch merger rate from dynamical channels. 
In Sec.~\ref{sec:discussion}, we discuss potential implications of our results, 
describing the  dependence on the PBH abundance and masses, and include an estimate for such a contribution in DM spikes consisting of PBHs  surrounding supermassive BHs (SMBHs).
We conclude in Sec.~\ref{sec:conclusions} with a discussion of potential future applications of the 3b scenario. 

\section{Binary formation in PBH clusters}\label{sec:binformc}

In the late Universe, after the epoch of recombination, new PBH binaries can only form dynamically through few-body processes. The close two-body encounter of two PBHs can induce a BPBH through a strong interaction. However, we also present an alternative binary formation mechanism below, via 3b interactions, which is most efficient in small clusters. To simplify the analysis, we take a monochromatic PBH mass spectrum fixed at a mass scale $m=30M_\odot$ and define $R_s=2Gm/c^2$
(where $G$ is the gravitational constant and $c$ is the speed of light)
as the Schwarzschild radius of the PBH.

\subsection{Two-body capture}
\label{sec.2body}

During the strong interaction of two compact objects, GW energy released at the point of closest approach may exceed the total energy of the system. Then, the energy of the two bodies becomes negative and a bounded pair forms as a consequence of energy conservation. The 
cross section for this capture mechanism to occur depends on the masses of the two bodies as well as their velocity dispersion $\sigma_v$.
It may be computed as follows~\cite{1989ApJ...343..725Q,Mouri:2002mc}:
\begin{align}
	\Sigma_\text{\tiny cap}&\simeq11R_s^2\left({c\over v_\text{\tiny rel}}\right)^{{18\over7}}\nonumber\\&\simeq1.2\times10^{-8}{\rm pc}^2\cdot\left({m\over30M_\odot}\right)^2\left({v_\text{\tiny rel}\over{\rm km/s}}\right)^{-{18\over7}}.
\end{align}
In this scenario, the two objects form a compact and highly eccentric binary which merges promptly, typically within only a few orbital cycles and 
with a maximum coalescence time of $\tau_\text{\tiny mrg}\approx3{\rm Myr}\cdot(m/(30M_\odot))(v_\text{\tiny rel}/(\rm km/s))^{-3}$~\cite{OLeary:2008myb}. Any PBHs which assemble to form a binary by capture at some redshift are assumed to directly merge at that redshift with an effectively negligible time delay.
Therefore, the binary formation rate can be translated into the merger rate for the capture scenario. The capture volumetric rate density is then calculated as
\begin{align}
	\gamma_\text{\tiny cap}\equiv n^2\langle\Sigma&_\text{\tiny cap}v_\text{\tiny rel}\rangle\simeq1.4\cdot10^{-5}{\rm Gyr}^{-1}{\rm pc}^{-3}\nonumber\\&\times\left({n\over{\rm pc}^{-3}}\right)^2\left({m\over30M_\odot}\right)^2\left({\sigma_v\over{\rm km/s}}\right)^{-{11\over7}},
	\label{eq.capRate}
\end{align}
where the angular brackets denote averages over the Maxwellian distribution for the relative velocity with parameter $\sqrt{2/3}\sigma_{v}$, and $n$ is the PBH number density.
The total rate per environment can be found by integrating this rate density over the volume of the cluster.
Evidently, capture rates are enhanced in denser systems with a  small velocity dispersion.

\subsection{Three-body interaction}
\label{sec.3body}

In the Newtonian regime, when the pericenter of interaction of two black holes is sufficiently larger than their horizons, GW emission is insufficient to induce a bound pair according to energy conservation. Nevertheless, 
the energy required to be released, for a bound system to be created, could be in the form of heat. That would be kinetic energy, absorbed by a third intermediary body which perturbs the two-body interaction in a short-lived 3b encounter. This energy extraction process becomes efficient in dense environments with relatively small velocity dispersion, so that gravitational focusing dominates the interaction and enhances binary formation from this channel.

Assuming that the 3b encounter occurs within a region of a given radius, the rate density for 3b encounters can be estimated as the product of the two-body interaction rate density $\gamma_{1,2}$ (taking into account both geometrical and gravitational focusing terms) 
times the probability that a third body happens to be in the same vicinity and participates in the interaction,
\begin{align}
p_{3,1-2}=\Gamma_{3,1-2}\tau_{1,2},
\end{align} 
where $\Gamma_{3,1-2}$ is the rate at which the third object encounters the interacting pair $1-2$ and  $\tau_{1,2}$ is the timescale of the two-body interaction~\cite{Ivanova:2005mi,2010ApJ...717..948I}. 
Moreover, we define the hardness ratio $\eta$ to be the binding energy of a binary with size $a$ normalized to the average kinetic energy of ambient single bodies, i.e.,
\begin{equation}
\eta\equiv \frac{Gm}{a\sigma_v^2}.
\end{equation}
Then, the rate for 3b encounters in a region of maximum size $a_\text{\tiny max}$ corresponds to a minimum threshold value $\eta_\text{\tiny min}$ for the hardness ratio.
Using this definition, the resulting rate density for three bodies to interact within that region can be expressed in terms of the minimum hardness ratio as~\cite{Rodriguez:2021qhl}
\begin{align}
\gamma_\text{\tiny 3b}(\eta\ge\eta_\text{\tiny min}) & =
{3^{9/2} \pi^{13/2} \over 2^{25/2}}
	\eta_\text{\tiny min}^{-{11\over2}}(1+2\eta_\text{\tiny min})\left(1+3\eta_\text{\tiny min}\right)
	\nonumber\\
	&\times{n^3(Gm)^5\over\sigma_v^9}.
	\label{eq:gamma3b}
\end{align}

The expression in Eq.~\eqref{eq:gamma3b} corresponds to the rate density for three single PBHs to interact and does not yet give us binary formation. It should be multiplied by the probability of binary formation by 3b encounters. This quantity was calculated numerically for equal masses
in Ref.~\cite{1976A&A....53..259A},
where it was found that if $\eta\gtrsim5$ then this probability asymptotically 
approaches
$100\%$. Therefore, for values of hardness ratio above 5, the 3b encounter rate essentially corresponds to the binary formation rate.
To account for the efficiency of binary formation in the hard region of the parameter space, we choose to set $\eta_\text{\tiny min}=5$ as used in the literature (see e.g. Refs.~ \cite{Morscher:2014doa,Rodriguez:2021qhl}). Using this value for the hardness ratio, the total BPBH creation rate density from the 3b channel becomes
\begin{align}
	\gamma_\text{\tiny 3b}(\eta\ge5)&\simeq3.8\cdot10^{-2}{\rm Gyr}^{-1}{\rm pc}^{-3}\nonumber\\&\times\left({n\over{\rm pc}^{-3}}\right)^{3}\left({m\over30M_\odot}\right)^5\left({\sigma_{v}\over{\rm km/s}}\right)^{-9}.
	\label{eq.3bRate}
\end{align}

For the 3b channel to matter at all, 3b encounters should be frequent.
PBHs that populate dense cluster environments have the chance to frequently interact among themselves and form 3b-induced binaries.
As for the capture channel, higher density environments with smaller velocity dispersion are preferred candidates where 3b interactions that induce hard binaries become important. Comparing Eq.~\eqref{eq.3bRate} with Eq.~\eqref{eq.capRate} in environments with $\sigma_v\approx1$~km/s and number density $n\approx1$~pc$^{-3}$, 3b binary formation is found to dominate over two-body capture. For example, PBH minihalos with similar characteristics are expected to form
naturally from the Poisson-induced PBH clustering at small scales. As we will describe in detail in Sec.~\ref{sec:mini-halos}, this process takes place at high redshift during the onset of the matter-dominated era.

\section{Fraction of coalescing binaries}
\label{sec:fraction}

Binaries formed by gravitational capture merge promptly with no substantial delay (Sec.~\ref{sec.2body}).
However, the 3b mechanism produces wide binaries which may merge with a significant delay that can exceed the age of the Universe. 
In this sense, the scenario proposed in this work resembles what happens to BPBH formation in the early Universe, in which close enough pairs of PBHs decouple from the Hubble flow and form an eccentric BPBH due to the torque from a third PBH in the vicinity \cite{Nakamura:1997sm,Ioka:1998nz}. 
In this section, we first determine the distribution of geometrical parameters describing BPBHs assembled via the 3b channel, and then compute the fraction of those binaries which merge within a predetermined time interval.

Equation~\eqref{eq:gamma3b} provides the formation rate density of permanent hard binaries via the 3b channel.
These binaries are hard in the sense that their binding energy $x\equiv Gm^2/(2a)$ (where $a$ indicates the binary's semimajor axis) is much greater than the average kinetic energy $m\sigma_v^2/2$ of a single object in the cluster. 
Since the 3b rate accounts for the formation of binaries with $\eta'$ larger than some threshold value $\eta$, Eq.~\eqref{eq:gamma3b} is proportional to the complementary cumulative distribution function for the hardness ratio. The probability density function (PDF) can be obtained by differentiating the negative of Eq.~\eqref{eq:gamma3b} with respect to $\eta_\text{\tiny min}$ and treating the result as a function of $\eta$ \cite{Morscher:2014doa}. In other words, we define
\begin{equation}
    P(\eta) \equiv
    \frac{1 }{\gamma_\text{\tiny 3b} (\eta_\text{\tiny min})}
    \left | 
    \frac{{\rm d}\gamma_\text{\tiny 3b} (\eta) }{{\rm d} \eta}
    \right|\,.
\end{equation}
If we limit ourselves to the range $\eta\ge5$ for the hardness ratio, as already discussed in the previous section, we can approximate the resulting PDF at leading order, i.e., $P(\eta)\propto\eta^{-{9/2}}$. 
We have checked that our results are only mildly sensitive to the exact value of the power-law index of this $\eta$ distribution. 
We discuss effect of dynamical binary-single hardening in the Appendix.

According to Refs.~\cite{1937AZh....14..207A,1975MNRAS.173..729H}, the distribution of eccentricities can be taken to be thermal, i.e. $P(e){\rm d}e={\rm d}e^2$, as long as the phase space density depends only on $x$, which is the case here.
For later convenience, we also make a change of variables and express the eccentricity in terms of $j=\sqrt{1-e^2}$, which is proportional to the angular momentum of the binary.
However, the statistical theory of resonant nonhierarchical triple encounters predicts that the distribution of eccentricities for the binary 
induced after the interaction could even be superthermal for low angular momentum encounters \cite{2019Natur.576..406S} (see also discussions in Refs.~\cite{Raidal:2018bbj,Vaskonen:2019jpv}). In this case, the distribution peaks at unity more prominently than the thermal distribution.
Subsequent interactions of the binary with other stars might thermalize their eccentricity, however, it is not clear if dynamical encounters can efficiently and fully thermalize an initially nonthermal distribution within the lifetime of a light cluster \cite{2019ApJ...872..165G}.\footnote{Notice that the thermal distribution is stationary and unaffected by dynamical relaxation, in the sense that even though encounters can alter the eccentricity of an individual binary, the average number density of binaries within a given eccentricity bin remains the same.}
Based on these considerations, in order to bracket uncertainties, we parametrize the normalized distribution of $j$ as a power law of the form
\begin{equation}
    P(j) = (1+\gamma ) j^\gamma,
\end{equation}
where $\gamma = 1$ for a thermal and $\gamma < 1$ for a superthermal distribution.
We later show that such a superthermal distribution can result in a significant enhancement in the merger rate, since a larger fraction of binaries is characterized by an eccentricity that allows mergers within a Hubble time.
On the other hand, the thermal distribution represents the conservative choice, providing a lower bound on the expected merger rate from 3b interactions.

\begin{figure}
	\centering
	\includegraphics[width=0.49\textwidth]{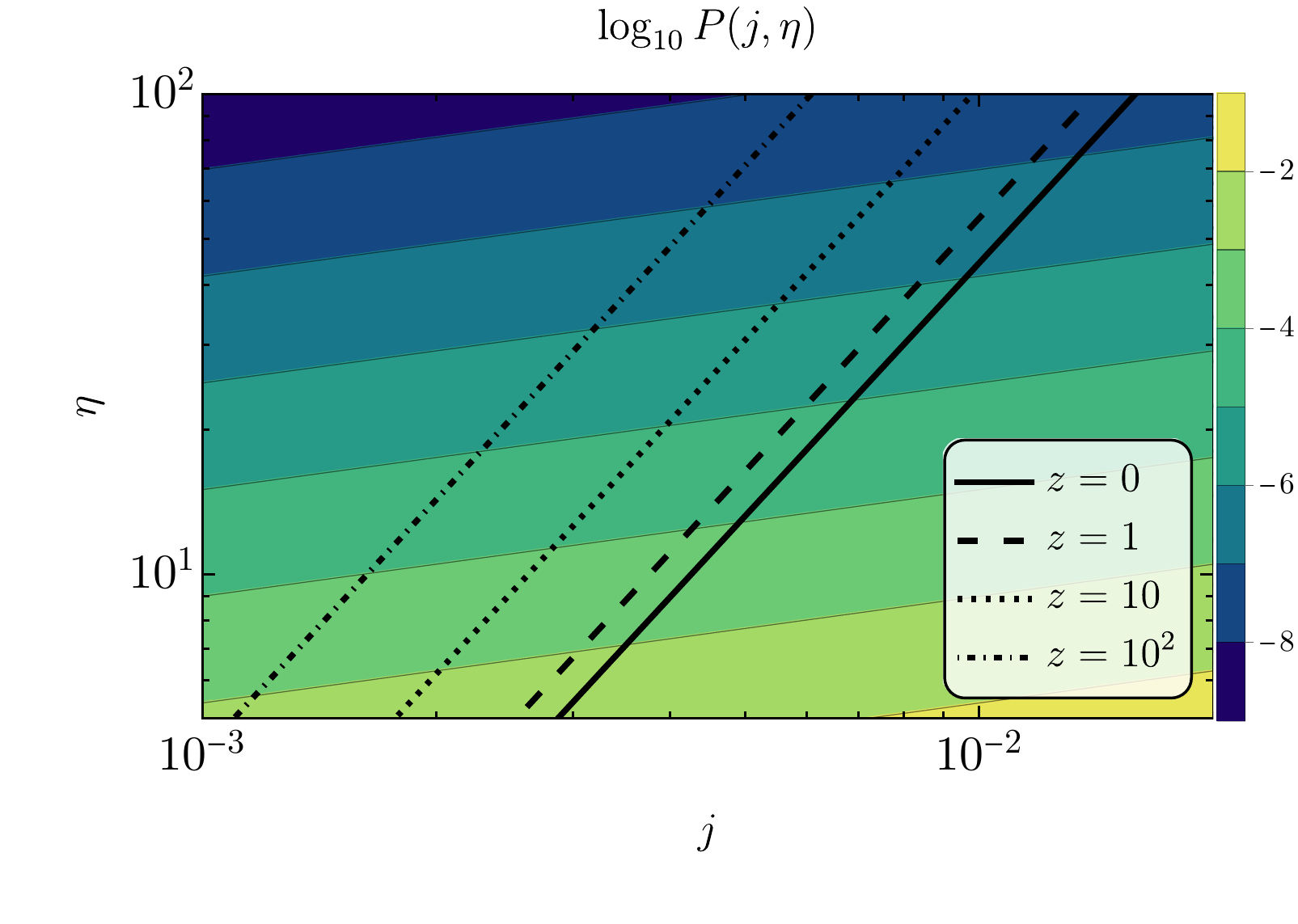}
	\caption{Combined distribution of $\eta$ and $j$ for binaries formed through 3b interactions. The black lines indicate the parameter space resulting in a merger time $t_\text{\tiny mrg}(j,\eta) = t(z)$, where $t(z)$ is the age of the Universe at redshift $z$.
	We assume $m_\text{\tiny PBH} = 30 M_\odot$, $\sigma_v = 1\,{\rm  km/s} $, and a thermal eccentricity distribution ($\gamma=1$).}
	\label{fig.dist_jeta}
\end{figure}

To summarize, the normalized joint distribution of binary size and shape is taken to be
\begin{align}
	P(j,\eta){\rm d}j{\rm d\eta}=
	\frac{7\cdot5^{7/2}}{2}(1+\gamma)
    j^\gamma
    \eta^{-9/2}
	{\rm d}j{\rm d\eta}
	\label{eq.jointPdf}
\end{align}
for $\eta\ge5$ and $j\in(0,1]$. 
The orbit of a newly formed BPBH can be tracked in the parameter space through the inspiral due to emission of gravitational radiation,
starting from the initial binding energy (or semimajor axis $a$) and angular momentum (or eccentricity $e$) which are controlled by $\eta$ and $j$, respectively.
The merger timescale can be computed in the high initial eccentricity  approximation as \cite{Peters:1963ux,1964PhRv..136.1224P}\footnote{
Reference~\cite{Mandel:2021fra} provides an accurate analytic fit of Peters's formula which is also valid for small eccentricities. We checked that this correction does not affect our results, as we are always well within the validity of the high initial eccentricity approximation.}
\begin{align}
	\tau_\text{\tiny mrg}(a,e)
	\simeq {3\over170}
	\frac{c^5a^4}{(Gm)^3} j^7.
	\label{eq.mergerTime}
\end{align}
 In Fig.~\ref{fig.dist_jeta} we plot the joint distribution $P(j,\eta)$  along with curves of constant $\tau_\text{\tiny mrg}(j,\eta)=T$.

To obtain the fraction of binaries that merge within a fractional time window $[T, T+\Delta T]$, 
we integrate the joint PDF in Eq.~\eqref{eq.jointPdf} with the right combination of $\eta$ and $j$ for which the merger time $\tau_\text{\tiny mrg}(j,\eta)$ is $T$, enforced through a Dirac delta function $\delta(\tau_\text{\tiny mrg}-T)$. 
This fraction becomes
\begin{align}
	Q(T)\equiv\int_0^1{\rm d}j \int_{\eta_\text{\tiny min}}^\infty{\rm d}\eta\ P(j,\eta) \delta(\tau_\text{\tiny mrg}(j,\eta)-T).
	\label{eq.def_Q}
\end{align}
We find that, since binaries assembled from 3b interactions regularly tend to be in the lower end of the hard region of the parameter space with $\eta\ge \eta_\text{\tiny min} = 5$ (bottom portion of the plot in Fig.~\ref{fig.dist_jeta}), 
it is required that they possess a large initial eccentricity to be able to merge within a Hubble time or less. 
As such, using Eq.~\eqref{eq.mergerTime}, we approximate the equation for the merger time as
\begin{align}
    &\tau_\text{\tiny mrg}
    (j\ll 1,\eta)
    \simeq{3\over340}{R_s\over c}\left({c\over\sigma_v}\right)^8{j^7\over\eta^4}
    \nonumber\\
    &\simeq
19\, {\rm Gyr}
\left({m\over30M_\odot}\right)
\left({\sigma_v\over{\rm km/s}}\right)
\left({j\over0.003}\right)^7\left({\eta\over5}\right)^{-4}.
\end{align}
To evaluate the integral in Eq.~\eqref{eq.def_Q}, we rewrite the Dirac delta as $\delta(\eta-\eta_0(j))$ times the appropriate Jacobian of the transformation. The symbol $\eta_0(j)$ corresponds to the hardness ratio as a function of angular momentum for which $\tau_\text{\tiny mrg}(j,\eta)=T$ has a solution.
The argument of the Dirac delta in Eq.~\eqref{eq.def_Q} has a solution as long as $j$ takes values above some minimum value $j_\text{\tiny min}$ and $\eta$ does not exceed a maximum value $\eta_\text{\tiny max}$. Physically, the former case corresponds to the widest binary we allow to form with $\eta=5$ which requires the maximum possible eccentricity (smallest $j$) to merge within $T$. The latter case corresponds to the tightest binary which merges in time $T$ if it starts with zero initial eccentricity. 
Notice that since the joint PDF is strongly tilted toward small values of $\eta$, the result of the integration is insensitive to the exact value of $\eta_\text{\tiny max}$. 
To find the minimum angular momentum,  we solve
the equation $\tau_\text{\tiny mrg}(j_\text{\tiny min},\eta=5)=T$ and obtain
\begin{align}
    & j_\text{\tiny min}\simeq
     5.0 
    \lp \frac{\sigma_v}{c} \rp^{\frac{8}{7}} 
    \left({ c T \over R_s}\right)^{1\over7}
    \nonumber\\
    &\simeq2.9\times10^{-3}\left({m\over30M_\odot}\right)^{-{1\over7}}\left({\sigma_v\over{\rm km/s}}\right)^{8\over7}\left({T\over13.8{\rm Gyr}}\right)^{1\over7}.
\end{align}
Therefore, the integral over $\eta$ can be performed first, and then we are left with the integration over $j$. This integration can be performed analytically in the low angular momentum approximation to get
\begin{align}\label{eq:Q}
    Q=
    \frac{7\cdot5^{7/2}(1+\gamma)}{2}
    \int_{j_\text{\tiny min}}^1
    {\rm d}j
    {j^\gamma\over(\eta_0(j))^{9\over2}}\left|{\partial\tau_\text{\tiny mrg}(j,\eta_0(j))\over\partial\eta}\right|^{-1}.
\end{align}
Finally, we find the probability to merge within $T$ per unit time
\begin{align}
 Q &  \simeq \frac{7 (1+\gamma)}{41- 8 \gamma}
 \lp \frac{212500}{3} \rp^{\frac{(1+\gamma)}{7}}
 \left({\sigma_v\over c}\right)^{8 (1+\gamma )\over7}
 \left({c T\over R_s}\right)^{(1+\gamma)\over7}
 {1\over T}.
\end{align}
Depending on the value of the angular momentum distribution exponent, one finds (fixing $m = 30 M_\odot$)
\begin{align}
    Q (\gamma=0)
    &  
    \simeq \frac{3.5\cdot10^{-5}}{\rm Gyr}
    \left({\sigma_v\over{\rm km/s}}\right)^{8\over7}\left({T\over13.8{\rm Gyr}}\right)^{-{6\over7}},
    \nonumber \\
    Q (\gamma=1)
    &  
    \simeq \frac{2.5\cdot10^{-7}}{\rm Gyr}
    \left({\sigma_v\over{\rm km/s}}\right)^{16\over7}\left({T\over13.8{\rm Gyr}}\right)^{-{5\over7}}.
\end{align}

As we will see in the following sections, the majority of the 3b-assembled BPBHs are efficiently formed in PBH minihalos with a relatively small number of members $N<10^2$.
As those environments quickly evaporate, those binaries essentially have to merge within a time window comparable to a Hubble time in order to be visible today. To give a back of the envelope estimate, if we assume that BPBH production is ongoing for $\approx22$~Myr at high redshift (which corresponds to the evaporation time of a cluster with 30 PBHs), then the probability that a binary formed via the 3b channel merges at the present epoch is $\approx4\cdot10^{-6}$. 
This means that only a few out of millions of PBH binaries assembled via 3b interactions at high redshift would be able to merge today.
Finally, as 3b binary formation is only effective at high redshift, the merger rate evolution observed at $z\lesssim {\cal O} (10)$ is dictated by the $Q$ factor alone. 

\section{$\text{3b}$ channel in PBH-induced small-scale structure}
\label{sec:mini-halos}

We now compute the contribution to the total PBH merger rate coming from binaries formed through 3b interactions in the PBH small-scale structure. 
We will consider the standard formation scenario, where PBHs 
follow a Poisson spatial distribution at formation \cite{Ali-Haimoud:2018dau,Desjacques:2018wuu,Ballesteros:2018swv,MoradinezhadDizgah:2019wjf,Inman:2019wvr,DeLuca:2020ioi}. 
We will first assume PBHs to be a large fraction of the DM abundance.
We will consider different environments and discuss how this result would scale with the PBH abundance $f_\PBH \equiv \rho_\text{\tiny PBH} / \rho_\text{\tiny DM}$ in the following section.

\subsection{PBH halo mass function}\label{sec:halomassfunction}
In this section, we analytically describe the small-scale structure induced by a population of PBHs dominating the DM budget formed with Poisson initial conditions (see e.g. Refs.~\cite{DeLuca:2020jug,Kadota:2020ahr}). 
This analytical description matches recent cosmological $N$-body simulations presented in Ref.~\cite{Inman:2019wvr}.
Models boosting the PBH correlation function at formation (e.g. with non-Gaussian curvature perturbations) 
are expected to enhance the formation of PBH small-scale structures, leading to higher 3b rates.
For this reason, the vanilla scenario we study here may be considered a conservative example of the relevance of the binary formation channel considered in this work. We will come back to this point in the conclusions. 

As the Universe evolves and structures form during the matter-dominated era, overdensities in the random field of PBHs at some point surpass the critical threshold for collapse $\delta_c\simeq1.686$ and decouple from the expansion to create virialized PBH minihalos. Depending on the number of objects $N$ in the cluster, this collapse occurs when the number variance ${\sigma}(N,z_f)=\delta_c$, where $z_f$ is the redshift of formation of a cluster with $N$ PBHs. 
Since on small scales Poisson perturbations dominate over adiabatic ones,
the characteristic density variance can be factorized into the product of the variance around the matter-radiation equality
\begin{equation}
    \sigma(N,z_\text{\tiny eq})\simeq1/\sqrt{N}
\end{equation}
and the growth factor \cite{Inman:2019wvr}
\begin{equation}
    D(z)=1+{3\over2}(1+z_\text{\tiny eq})/(1+z)
\end{equation}
describing the evolution of matter perturbations, where $z_\text{\tiny eq} =3402$ is the redshift at the matter-radiation equality. Thus, the condition for collapse translates into an equation for the formation redshift of PBH minihalos as a function of number $N$ as 
\begin{align}
    z_f={3\over2}\frac{(1+z_\text{\tiny eq})}{\delta_c\sqrt{N}-1}-1\simeq0.890{z_\text{\tiny eq}\over\sqrt{N}}.
\end{align}
When virialized, and in the approximation of top-hat collapse, newly born PBH minihalos have an average density given by ${\rho}_\text{\tiny cl}\simeq200\cdot \overline\rho_\text{\tiny cr}(z_f)$ in terms of the critical density of the Universe $\overline \rho_\text{\tiny cr}(z_f)$ evaluated at the redshift of cluster formation. The symbol $M=Nm$ denotes the mass of the cluster with $N$ members, and the size of the system $R$ is determined by the condition 
${4\over3}\pi R^3{\rho}_\text{\tiny cl}=M$. 
The characteristic velocity dispersion $\sigma_v$ is then evaluated by applying the virial theorem as $\sigma_v^2=0.8GM/R$ \cite{1971ApJ...164..399S}. 

One can describe the distribution of halos formed from the collapse of increasingly large overdense regions adopting the Press-Schechter theory \cite{1974ApJ...187..425P}. 
The differential comoving number density of clusters with $N$ objects is found to be
\begin{align}\label{eq.haloMF}
	{{\rm d} n_\text{\tiny cl}(N,t)\over {\rm d} N}={\overline{n}\over\sqrt{\pi}}
	\left[{N\over N_{*}(t)}\right]^{-{1\over2}}
	{e^{-N/N_{*}(t)}\over N^2}\,,
\end{align}
where we introduced the mean number of PBHs per unit comoving volume \cite{Sasaki:2018dmp}
\begin{equation}
\overline{n} \equiv f_\PBH \frac{\rho_\text{\tiny DM} }{m} 
= 1.1 \,{\rm kpc}^{-3} f_\PBH \lp { m \over 30 M_\odot}\rp^{-1}. 
\end{equation}
The characteristic halo size $N_*(t)$ is instead fixed by evaluating the
number of objects whose Poisson perturbations ($\approx 1/\sqrt{N}$) are able to meet the threshold at the given epoch and turns out to be \cite{Hutsi:2019hlw,DeLuca:2020jug}
\be
\label{dd}
N_*(t) \simeq f_\PBH^2\left(\frac{2600}{1+z}\right)^2.
\ee

It is important to stress that small PBH clusters are characterized by a finite lifespan. 
Indeed, internal evolution of the cluster via two-body relaxation causes the evaporation of PBHs from the system until the minihalo dissolves completely or is engulfed in a larger halo. 
The lifetime of minihalos is characterized by $t_\text{\tiny ev}\simeq140t_\text{\tiny rlx}$ where the relaxation timescale is (e.g. \cite{1987gady.book.....B})
\begin{equation}
    t_\text{\tiny rlx} \simeq 
    {1 \over 10}
    \frac{N}{\ln N}
    \left ( \frac{R}{\sigma_{v}} \right ).
\end{equation}
Therefore the evaporation time is given by 
	\begin{equation}
	t_\text{\tiny ev}
	\simeq \frac{1.4 {\rm Gyr} }{\ln N} 
	\left( \frac{N}{100} \right)^{1/2} 
	\left( \frac{m}{30 M_\odot}\right)^{-1/2}
	\left( \frac{R}{{\rm pc}} \right)^{3/2}
	\end{equation}
as a function of the typical cluster virialization radius $R$.
We derived an accurate fit of the size of the cluster which is expected to evaporate at redshift $z_\text{\tiny ev}$. This takes the form 
\begin{equation}\label{Nevapfit}
    N_\text{\tiny ev}(z_\text{\tiny ev})
    =
    \frac{2190}{\lp1+z_\text{\tiny ev}\rp^{0.9734}}
    -
    \frac{526.5}{\lp1+z_\text{\tiny ev}\rp^{1.909}} 
\end{equation}
and is valid for redshifts in the range $z_\text{\tiny ev}\in [0, 10^3]$. 

\begin{figure}
	\centering
	\includegraphics[width=0.49\textwidth]{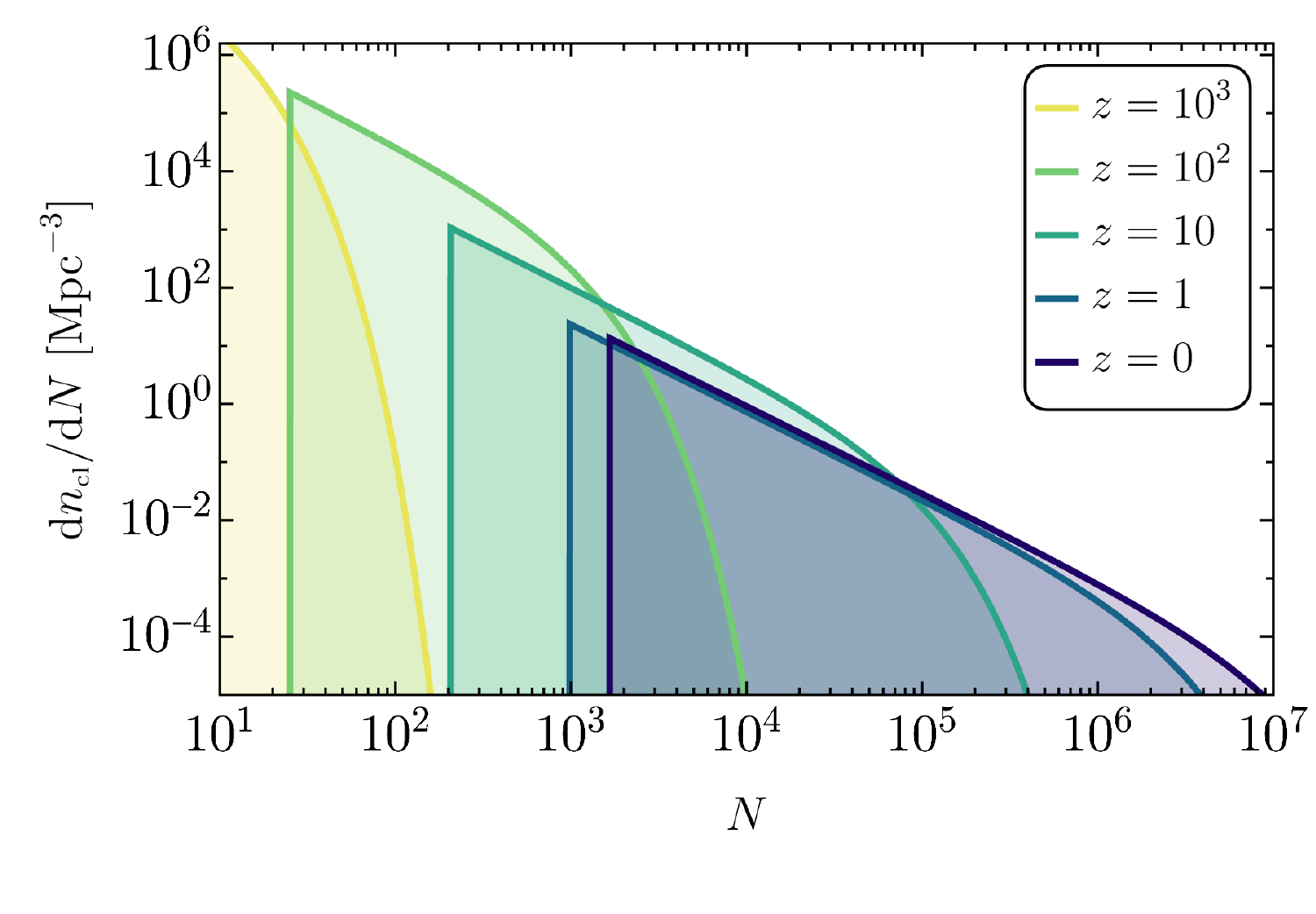}
	\caption{ 
	Number density of PBH minihalos $dn_\text{\tiny cl}/dN$ in units of comoving volume (Mpc$^{-3}$) as a function of the number $N$ of PBH members in the cluster for a set of redshift values. The hard cutoff in the left part of the halo mass function comes from the fact that smaller halos that formed at even larger redshift have evaporated by the observation redshift.
	This plot neglects adiabatic perturbations responsible for large-scale structure development at low redshift.}
	\label{fig.Halos}
\end{figure}

As structure formation proceed hierarchically from the  bottom up, 
there is also a nonvanishing probability of larger halos engulfing smaller PBH clusters.
The finite lifespan of small halos is, therefore, dictated by both evaporation timescale and survival probability (see Ref.~\cite{DeLuca:2020jug} and references therein). 
We include these effects in the computation of the halo mass function by accounting for the time evolution of ${\rm d} n_\text{\tiny cl}/{\rm d} N$ and cutting the contribution 
from clusters smaller than $N_\text{\tiny ev}(t)$, which is the size of the clusters whose evaporation time is $t_\text{\tiny ev} = t$. 
In other words, we write
\begin{equation}
  {{\rm d} n_\text{\tiny cl}^\text{\tiny ev}(N,t)\over {\rm d} N}
 =
 {{\rm d} n_\text{\tiny cl}(N,t)\over {\rm d} N} 
 \times \Theta(N - N_\text{\tiny ev}(z))\,,
\end{equation}
where $\Theta$ is the Heaviside function.
We plot the halo mass distribution at various epochs in Fig.~\ref{fig.Halos}.

\begin{figure*}
	\centering
	\includegraphics[width=0.49\textwidth]{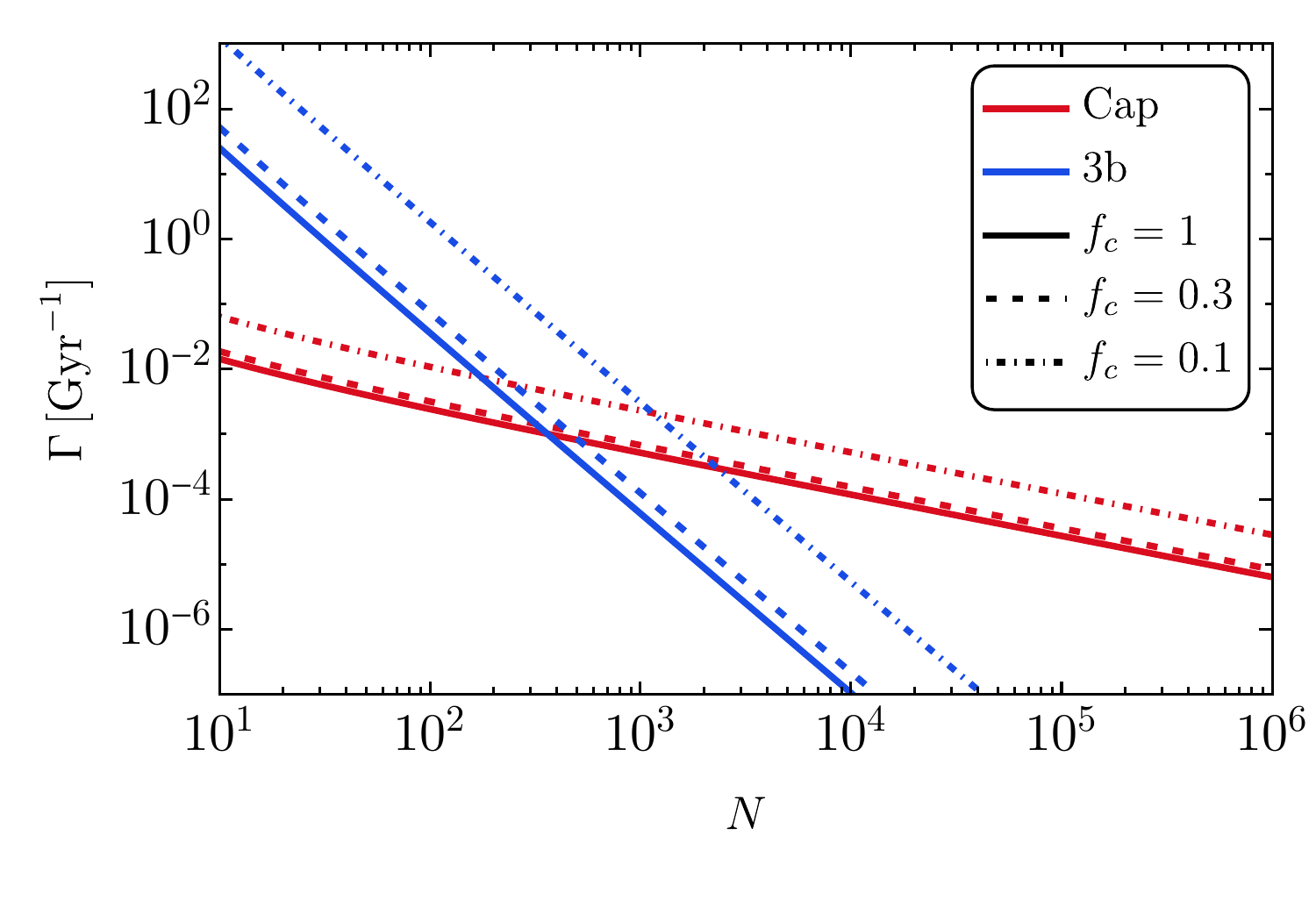}
	\includegraphics[width=0.49\textwidth]{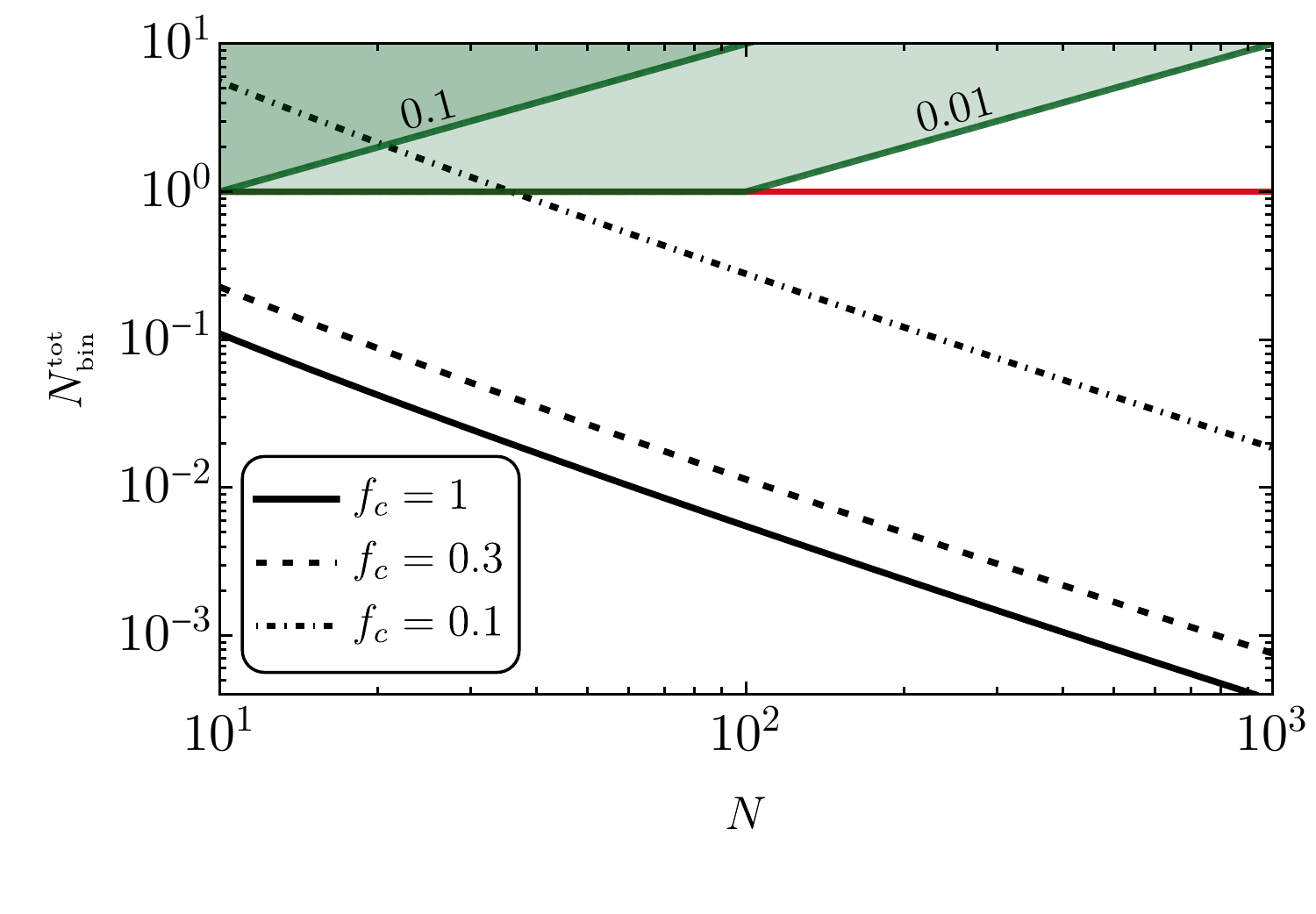}
	\caption{ 
	{\bf Left: }
	binary formation rate per cluster for the 3b and capture channels as a function of the number of cluster members. We consider three different values for $f_c$ as indicated in the legend.  
	Note that BPBH formation rate in the capture case, unlike for 3b, roughly equals the merger rate, because captured pairs coalesce promptly with negligible time delays.
{\bf 	Right:}
total number of binaries $N^\text{\tiny tot}_\text{\tiny bin}$ formed in each environment with $N$ objects in a timescale comparable to the evaporation time. 
The red (green) line delimits the region above which more than 1 ($N_\text{\tiny tot}^\text{\tiny bin}/N= 0.01,0.1$) binaries are formed within the cluster lifetime. }
	\label{fig.Rates&Halos}
\end{figure*}

\subsection{PBH cluster properties}
The rate of dynamical interactions crucially depends on the number density of PBHs and their characteristic relative velocity. Therefore, it is important to include modeling of the cluster density profiles in the estimates for the rate. 

Here we follow the analytical description of the PBH cluster profiles derived in Ref.~\cite{DeLuca:2020jug}, 
which is consistent with numerical simulations of single PBH cluster dynamics in Ref.~\cite{Raidal:2018bbj}.
PBH clusters are characterized by a density profile scaling like $\rho \propto r^{-12/5}$. It is important, however, to include the formation of an inner core induced by gravitational interactions between the objects composing the cluster. 
In order to bracket uncertainties, we are going to parametrize the size of the core $r_c$ as a function of the fraction of objects $f_c$ residing within $r<r_c$. 
In other words, we assume
\begin{align}
	\rho(r)=\begin{cases}
		\rho_c&,\quad r<r_c, \\
		\rho_c\left({r\over r_c}\right)^{-{12\over5}}&,\quad r\in[r_c,R].
	\end{cases}
\end{align}
As a consequence, the total mass $M$, i.e., the integral of the density profile over $[0,R]$, 
sets the normalization of the core density $\rho_c$, 
while the core is fixed by the fraction of mass within $r_c$ as $M_c = f_c M$.
 We also fix the velocity to the virialized velocity of the cluster, which means
\begin{equation}
\sigma_v^2=\frac{4GM}{5R}.
\end{equation}
We checked that allowing for the mean velocity to vary with radius would not affect our result. 
In such a case, adopting a parametrization of the velocity of the form $\sim{4GM(r)}/(5r)$, one would observe a small velocity reduction near the core center. In this sense, our choice is conservative as it leads 
to estimating a smaller rate. 

To get a reasonable estimate for the fraction of PBHs residing in the core $f_c$, we compare with profiles of star clusters in astrophysical environments.
In those systems, the Plummer sphere \cite{1911MNRAS..71..460P} provides a reasonable model for the radial distribution of stars in globular clusters. Integrating this profile up to the core radius, one finds that $f_c\approx16\%$ of the total mass is contained within the core of the cluster. 
Lacking numerical simulations of small PBH halos at scales close to the center of the cluster, we consider three distinct values for  $f_c\in\{0.1,0.3,1\}$.
Notice that $f_c=1$ corresponds to a constant density profile, where density and velocity are fixed by the virial mass and radius of the cluster.
This assumption is often adopted in the literature, and it provides a conservative lower bound on the 3b merger rate.

In the left panel of Fig.~\ref{fig.Rates&Halos} we show the binary formation rates $\Gamma$ computed in clusters of $N$ objects for different choices of $f_c$. As one can see, the rate of binary formation from the 3b channel drastically 
exceeds the one from   dynamical capture 
when $N\lesssim 10^3$. 
It is important to notice, however, that merger time delays for 3b binaries are much higher, a property which is going to decrease the difference between the contributions from the two channels. 

In the right panel of Fig.~\ref{fig.Rates&Halos} we report the total number of binaries formed in a cluster lifetime, $N_\text{\tiny bin}^\text{\tiny tot} =  t_\text{\tiny ev} \times \Gamma$.
As the formation of a binary within a cluster composed of a small number of objects would potentially affect its properties (such as density profile and velocity dispersion), we conservatively cut the total number of binaries formed within a cluster to unity.
This is expected to affect only the case of clusters with a small core and large central density, i.e. $f_c=0.1$, for which the rate is sufficiently high. 

\subsection{Merger rate of dynamically formed binaries}\label{sec:mergerrate}

Having estimated the rate of binary formation per cluster, we are left with the integration over the number density of clusters expected to be present in the Poisson-induced small-scale structure. 
Considering first the rate of binaries from the capture channel at present time, we compute
\begin{align}
	{\cal R}^\text{\tiny cap}_\text{\tiny BPBH}(z)
	&=
\sum_{N=N_\text{\tiny ev}(z)}^{N_*(z)}
	\Gamma_\text{\tiny cap}(N){dn_\text{\tiny cl}^\text{\tiny ev}(N,z)\over dN}.
\end{align}
We iterate that this estimate assumes negligible time delays compared to $t(z)$.
Notice that the summation over the cluster size $N$ only starts from the smallest clusters which have not evaporated yet at redshift $z$, see Eq.~\eqref{Nevapfit}.
In Table~\ref{Tb:parameters} we report the corresponding capture rates in the local Universe ($z=0$).
The sum is dominated by contributions coming from the smallest clusters close to $N_\text{\tiny ev}(z)$, for which the rate is higher (see Fig.~\ref{fig.Rates&Halos}) and the cluster number density peaks.

In the estimate of the merger rate from capture, we neglect the impact of adiabatic perturbations, which would reduce the fraction of mass residing in small-scale structures in the late-time Universe
by boosting the collapse and virialization of structures above galactic scales.
Including this effect would move a larger fraction of DM into virialized structures of much larger sizes (with corresponding larger velocity dispersion) and smaller densities, where the dynamical formation of binaries is quenched.
While this may suppress the rate of binaries formed in the late-time Universe ${\cal R}^\text{\tiny cap}_\text{\tiny BPBH}$, it does not affect the 3b merger rate, which is dominated by binaries formed in very small clusters which are evaporating at redshift larger than ${\cal O}(10-100)$.

{
\renewcommand{\arraystretch}{1.4}
\setlength{\tabcolsep}{10pt}
\begin{table}
\caption{Merger rate density of dynamically formed binaries at redshift $z\simeq 0$ in units of $({\rm yr^{-1}Gpc^{-3}})$ for capture and 3b channels and assuming $m=30M_\odot$.
We assume either a thermal $(\gamma=1)$ or  superthermal $(\gamma=0)$ eccentricity distribution.
We conservatively integrate the rate from $N_\text{\tiny min} = 10$ and require at most the formation of one binary per cluster 
(only affecting the 3b rate for  $f_c=0.1$).}
\vspace{.1cm}
\begin{tabular}{cccc}
\hline
\hline
$f_c$ &  
${\cal R}^\text{\tiny cap}_\text{\tiny BPBH}$& 
${\cal R}^\text{\tiny 3b}_\text{\tiny BPBH}(\gamma=1)$ &
${\cal R}^\text{\tiny 3b}_\text{\tiny BPBH}(\gamma=0)$
\\
\hline
\hline
1 & 7.3 & 2.7 & $1.3 \times 10^2$ \\
\hline
0.3 & 9.6 & 5.5 & $2.7 \times 10^2$ \\
\hline
0.1 & $3.3 \times 10^1$ & $5.1 \times 10^1$ & $2.5 \times 10^3$ \\
\hline
\hline
\end{tabular}
\label{Tb:parameters}
\end{table}
}

We compute the merger rate density of binaries produced by 3b interactions by integrating the binary formation rate over the age of the Universe and by multiplying by the fraction of binaries merging within the remaining time window $t(z)-t'$
using Eq.~\eqref{eq:Q}, summed over the halo mass function. 
Therefore, we compute
\begin{align}\label{3bratetoday}
	{\cal R}^\text{\tiny 3b}_\text{\tiny BPBH}(z)&=
\sum_{N=N_\text{\tiny min}}^{N_*(z)}
\Bigg [
\Gamma_\text{\tiny 3b}(N)
\nonumber \\
&\times
\int_{t\text{\tiny min}}^{t(z)}
\d t' 
Q(N,t(z)-t'){\d n^\text{\tiny ev}_\text{\tiny cl}(N,t')\over  \d N}
\Bigg ],
\end{align}
where $t(z)$ is the age of the Universe at redshift $z$ and $t\text{\tiny min}\equiv t_{f}(N_\text{\tiny min})$. This integral already accounts for the cluster evaporation timescale through the halo mass function $\d n^\text{\tiny ev}_\text{\tiny cl}/\d N$.
In Table~\ref{Tb:parameters} we report the 3b rate obtained for $z\simeq 0$ for different values of $f_c$
and two assumptions on the eccentricity distribution of binaries, following
either a thermal $(\gamma=1)$ or superthermal $(\gamma = 0)$ distribution. 
Notice that the current 3b merger rate density is comparable to the capture rate if one assumes $\gamma = 1$, while it becomes ${\cal O}(10^2)$ times larger in case of a superthermal distribution. 

We conservatively report results integrating from clusters larger than $N_\text{\tiny min} = 10$.
Including even smaller clusters in the count would boost the estimated 3b rate due to the larger number density of small clusters and higher rates obtained in those environments (see Fig.~\ref{fig.Rates&Halos}). However, the dynamics of such small clusters may deviate from the modeling discussed above and should be estimated with dedicated few-body simulations. 

\section{Discussion}\label{sec:discussion}

In the previous section we presented the computation of the merger rate from 3b interactions by assuming a large value of the abundance $f_\PBH$, showing the potential relevance of  this channel, largely neglected in the PBH literature. 
In this section, we discuss the implications 
for various PBH scenarios. 
With this aim, we start by comparing our results to the merger rate of binaries produced in the early Universe. 

\subsection{Comparison with the merger rate of binaries formed in the early Universe}
PBH binaries can form in the early Universe out of decoupling from the Hubble flow before  matter-radiation equality~\cite{Nakamura:1997sm,Ioka:1998nz}. 
Assuming a narrow mass distribution, the differential volumetric PBH merger rate density takes the form~\cite{Raidal:2018bbj, Vaskonen:2019jpv,DeLuca:2020jug,DeLuca:2020qqa}
\begin{align}
\label{eq:diffaccrate}
 {\cal R}^\text{\tiny EU}_\text{\tiny BPBH}(z)
& = 
\frac{7.1 \times 10^2}{{\rm Gpc^3 \, yr}} 
f_\PBH^{\frac{53}{37}} 
\lp \frac{t(z)}{t_0} \rp^{-\frac{34}{37}}  
\lp \frac{m}{ 30 M_\odot} \rp^{-\frac{32}{37}}  
 \nonumber \\
& \times
\left[ \frac{S(m, f_\PBH,t(z))}{2.4 \cdot 10^{-3}} \right],
\end{align}
where the suppression factor $S<1$  accounts for environmental effects in both the early- and late-time Universe, 
normalized to 
its value when $f_\PBH = 1$ and $z=0$.
The effects suppressing the early Universe merger rate contained in $S$ can be divided in two categories. 
In the early Universe, close to the binary formation epoch, this accounts for interactions between PBH binaries and both surrounding DM inhomogeneities and neighboring isolated PBHs~\cite{Eroshenko:2016hmn,Ali-Haimoud:2017rtz,Raidal:2018bbj,Liu:2018ess}. 
In the late Universe, this includes the successive disruption of binaries that populate PBH clusters formed from the initial Poisson conditions~\cite{Vaskonen:2019jpv,Jedamzik:2020ypm,Young:2020scc,Jedamzik:2020omx,DeLuca:2020jug,Trashorras:2020mwn,Tkachev:2020uin,Hutsi:2020sol,link} throughout the evolution of the Universe.\footnote{In this context, we define a disrupted binary as one whose semimajor axis and eccentricity are modified following a binary-single interaction. 
As found in the $N$-body simulation of Ref.~\cite{Jedamzik:2020ypm}, such events in small PBH clusters tend to circularize the orbits and enhance the merger time delays.}
An analytic expression for $S$ can be found in Ref.~\cite{Hutsi:2020sol}.
The combination of abundance-dependent factors in Eq.~\eqref{eq:diffaccrate} results in an effective scaling of the rate proportional to
\begin{align}
	{\cal R}^\text{\tiny EU}_\text{\tiny BPBH}
	\propto
	\begin{cases}
		f_\PBH^{2/3} &,\quad f_\PBH\gtrsim 10^{-3},
		\\
		f_\PBH^{2} &,\quad f_\PBH \lesssim 10^{-3}.
	\end{cases}
\end{align}
We are not including here the contribution to the merger rate from initial PBH binaries which are disrupted in PBH clusters, which may still be sizeable for values of the abundance close to unity~\cite{Vaskonen:2019jpv}.

It is interesting to mention that, even though predictions for low-redshift observables are the same as for binaries formed in the early Universe, i.e., the eccentricity is lost by GW emission before detection \cite{Franciolini:2021xbq}
and accretion effects would induce mass-spin correlations below redshift $z\lesssim 30$ (see e.g. Refs.~\cite{DeLuca:2020fpg,DeLuca:2020bjf,Franciolini:2022iaa}), 
the 3b channel predicts a different redshift evolution of the merger rate compared to the early Universe,  scaling as ${\cal R}^\text{\tiny 3b}_\text{\tiny BPBH}\approx  t^{(\gamma-6)/7}$.
Therefore, this channel can be, in principle, distinguishable from the other contributions. 

By comparing the PBH binary merger rate of dynamical channels reported in Table~\ref{Tb:parameters} and the 
early Universe contribution, 
we see that the latter always dominates the overall merger rate, unless PBHs made up a dominant fraction of the dark matter above the solar mass range, a scenario which is ruled out by current constraints \cite{Carr:2020gox},
and the 3b channel is characterized by the superthermal distribution and $f_c = 0.1$.
In such a case, the 3b rate alone would be too large to be compatible with the rate of binary BHs observed by the LVKC
for objects around $\approx 30 M_\odot$~\cite{2021arXiv211103634T}. 
Therefore, our results both strengthen the LVKC bound dictating that PBHs of tens of solar masses cannot comprise all of the DM and confirm the merger rate of binaries formed in the early Universe is the dominant channel in the standard PBH formation scenario.

To reduce the uncertainties affecting the computation of the 3b channel, we need further numerical investigations of cluster profile evolution and eccentricity distribution attained in 3b binary formation.
Even in the most conservative estimates, assuming a thermal distribution of eccentricity and boxlike PBH clusters (i.e., $f_c=1$),
the rate is still comparable to that from dynamical capture and thus should be considered in the computation of PBH rates when different scenarios are explored. 
In the following subsection, we are going to describe how our result would change by varying the PBH mass, abundance, and environment.

\subsection{Scaling with the PBH abundance}

In the preceding discussion, we always assumed PBHs contribute to a dominant fraction of the DM.
Let us derive the expected scaling of this rate as a function of $f_\text{\tiny PBH}$.

Various differences are expected when decreasing $f_\text{\tiny PBH}$ below unity. The first effect is that the PBH cluster mass function directly scales with the PBH number density $\bar n \rightarrow f_\text{\tiny PBH} \bar n$.
Additionally, Poisson perturbations induced by the PBH population do not involve the secondary DM component. That means the threshold for cluster collapse can be scaled as $\delta_c \rightarrow \delta_c/f_\text{\tiny PBH}$. 
As a consequence, the growth factor is required to make up for the increased effective threshold for collapse, decreasing the formation redshift of PBH clusters. 
One finds $z_f \rightarrow z_f f_\text{\tiny PBH}$, under the assumption of $D(a) \sim a$ in the relevant redshift range. 
As PBH clusters form later in the evolution of the Universe, their virial density becomes $\rho_\text{\tiny cl} \rightarrow \rho_\text{\tiny cl} f_\text{\tiny PBH}^3$ and the size scales as $R\rightarrow R/ f_\text{\tiny PBH}$. Also, as a consequence, the velocity dispersion scales as $\sigma_v \sim \sqrt{M/R} \sim f_\text{\tiny PBH}^{1/2}$. 
Finally, the fraction of binaries merging within a given time interval close to the present epoch roughly scales as 
$Q\sim \sigma_v^{1+\gamma} \sim f_\PBH^{(1+\gamma)/2}$.
Accounting for these effects, we can roughly expect the rate to scale as 
\begin{equation}\label{scale_3b}
    {\cal R}^\text{\tiny 3b}_\text{\tiny BPBH}\propto  
    Q
    \times 
    R^3
    \bar n 
    \times
    t_\text{\tiny ev}
    \frac{n^3}{\sigma_v^9}
    \propto
    f_\text{\tiny PBH}^{2+(1+\gamma)/2}.
\end{equation}
This scaling is derived by  accounting for the larger evaporation time of PBH clusters obtained for smaller $f_\text{\tiny PBH}$, i.e. $t_\text{\tiny ev} \sim R/\sigma_v \sim  f_\text{\tiny PBH}^{-1/2}$.
With a similar estimate, we find that the merger rate from capture scales as 
\begin{equation}\label{scale_cap}
    {\cal R}^\text{\tiny cap}_\text{\tiny BPBH}(z)
    \propto
     R^3 \bar n
    \times
    \frac{n^2}{\sigma_v^{11/7}}
    \propto f_\text{\tiny PBH}^{3},
\end{equation}
which is faster than the 3b rate.
It is worth mentioning that the scaling derived above does not include the potential effect of PBH segregation in mixed DM clusters, which may induce a boost of rates for dynamically formed binaries. 

If we compare the contribution from dynamical channels to that from early Universe binaries, which scales as ${\cal R}^\text{\tiny EU}_\text{\tiny BPBH} \approx f_\PBH^{2/3}$ for large values of the abundance, we see that the latter becomes increasingly dominant when $f_\PBH$ becomes smaller and smaller. 

\subsection{Asteroid mass PBHs}

Another interesting mass range for PBHs is the so-called asteroid mass range, which approximately spans $m\in [10^{-16}, 10^{-10}]$, where there are no constraints on the PBH abundance \cite{Katz:2018zrn,Montero-Camacho:2019jte}.
The frequency of GWs emitted from such light BH mergers is related to the innermost stable circular orbit (ISCO) frequency by
 \begin{equation}
\label{eq:fISCO}
f_\text{\tiny ISCO} \simeq  4.4 \times 10^3 \, {\rm Hz} 
\lp \frac{m_1 + m_2}{M_{\odot}} \rp^{-1},
\end{equation} 
where $m_1$ and $m_2$ are the masses of the two BHs.
Therefore, we immediately see that such light mergers would produce GW signals outside the detectability band of ground- and space-based interferometers, but could be the target of ultrahigh frequency GW searches (see Ref.~\cite{Franciolini:2022htd} and references therein).

Here we estimate how the merger rate of 3b binaries scales with the PBH mass.
As the number density of PBH clusters scales proportionally to the PBH number density [see Eq.~\eqref{eq.haloMF}], one finds that it scales as
$\bar n \sim m^{-1}$.
Additionally, the collapsed PBH clusters are characterized by a density roughly $200$ times the mean density in the Universe at cluster formation (that does not depend on PBH masses), the size of clusters scales like 
$R \sim  (N m / \rho_\text{\tiny cl})^{1/3}\sim m^{1/3}$,
and the virial velocity (i.e., approximately the characteristic PBH relative velocity) is 
$\sigma_v \sim (N m  /R)^{1/2} \sim m^{1/3}$.
As a consequence, the cluster evaporation time $t_\text{\tiny ev}$
becomes independent of the PBH masses. 
Finally, from Eq.~\eqref{eq:Q}, we find that the $Q$ factor scales as $Q \sim m^{5(1+\gamma)/21}$.
Collecting all contributions, one obtains
\begin{equation}
    {\cal R}^\text{\tiny 3b}_\text{\tiny BPBH}\propto  
    Q
    \times 
    R^3
    \bar n 
    \times 
    t_\text{\tiny ev}
    \frac{m^5 n^3}{\sigma_v^9}
    \propto
   m^{-1 + 5 (1 + \gamma)/21}.
\end{equation}
Depending on which distribution of eccentricity is assumed, this becomes either $\propto m^{-11/21}$ for $\gamma=1$ or $\propto m^{-16/21}$ for $\gamma=0$. 
Using a similar procedure, the scaling of the capture channel is found to be
\begin{equation}
    {\cal R}^\text{\tiny cap}_\text{\tiny BPBH}(z)
    \propto
    R^3  \bar n
    \times 
    \frac{m^2 n^2}{\sigma_v^{11/7}}
    \propto
    m_\text{\tiny PBH}^{-11/21},
\end{equation}
which is, strikingly, the same scaling found for the 3b channels with the thermal distribution.  

However, both contributions are largely subdominant for small PBH masses compared to the early Universe merger rate, which scales as ${\cal R}^\text{\tiny EU}_\text{\tiny BPBH} \propto m^{-32/37}$.
We conclude that both dynamical channels are subdominant as far as the asteroid mass range is concerned. 

\subsection{Dark matter spikes}

PBHs act like cold DM and generically form density spikes around SMBHs
\cite{Gondolo:1999ef,Bertone:2005xv,Ferrer:2017xwm}. The spike density profile is sensitive to the dynamical history of the SMBH and varies between $9/4$ and $3/2$. Most analyses converge on a $7/3$ profile expected for an ambient Navarro-Frenk-White halo profile, due to the likely sparsity of late merging events. 
The high spike density may boost PBH merger rates, and here we evaluate the contribution from the 3b channel.

PBHs may sink into the central density spike by the action of dynamical friction. For nearly circular PBH orbits of radius $r$, the Chandrasekhar expression for dynamical friction on PBHs of mass $m$ in a predominantly cold DM spike is $t_\text{\tiny df}(r)/t_{\rm circ}\approx C_\text{\tiny df}M(<r)/m$, where $C_\text{\tiny df}$ contains a logarithmic term and is approximately of order 10. We adopt a $7/3$ spike and set the spike radius to
\begin{align}
    r_\text{\tiny sp}& ={GM_\text{\tiny SMBH}\over\sigma_\text{\tiny sp}^2}
    \nonumber \\
    &\simeq200{\rm pc}\left({M_\text{\tiny SMBH}\over6.5\times10^9M_\odot}\right)\left({\sigma_\text{\tiny sp}\over{\rm 400km/s}}\right)^{-2},
\end{align}
where $\sigma_\text{\tiny sp}$ is the velocity dispersion at $r=r_\text{\tiny sp}$. We checked that, in the vicinity of the SMBH for $r<r_\text{\tiny sp}$, the contribution of the spike to the mass enclosed can be neglected. Thus, we compute the velocity dispersion as $\sigma^2(r)=GM_\text{\tiny SMBH}/r$.

Taking as an example the DM halo of our Milky Way, we integrate the merger rate density for the capture and 3b channels over radius from $8GM_\text{\tiny SMBH}/c^2$ up to $r_\text{\tiny sp}$ following the computation performed in Ref.~\cite{Nishikawa:2017chy}. 
We find $\Gamma_\text{\tiny sp}^\text{\tiny cap}\approx10^{-9}$yr$^{-1}$ for the capture merger rate contribution from the DM spike of a Milky Way-like galaxy. However, the contribution from the 3b channel lies 23 orders of magnitude below the capture counterpart, which indicates that 3b interactions would not contribute to the dynamical formation of BPBHs in those environments. 
The reason for this large suppression of the 3b channel 
lies in the fact that the PBH velocity dispersion is dominated by the gravitational potential of the central SMBH and is very large. Since the 3b rate depends on the velocity dispersion through the  factor $\gamma_\text{\tiny 3b} \propto n^3 \sigma^{-9}$, 
the enhanced high central density is not able to 
compensate this trend.
We conclude that 3b interactions do not represent an efficient binary formation channel for DM spikes surrounding SMBHs.

\section{Conclusions}\label{sec:conclusions}

We investigated the PBH binary merger rate resulting from dynamical
scenarios. By adopting well-known results in the astrophysical context to describe 3b interaction rates in star clusters, we analytically computed the rate of 3b binary formation and merger time delays in the PBH-induced small-scale structure.

\begin{figure}
	\centering
	\includegraphics[width=0.49\textwidth]{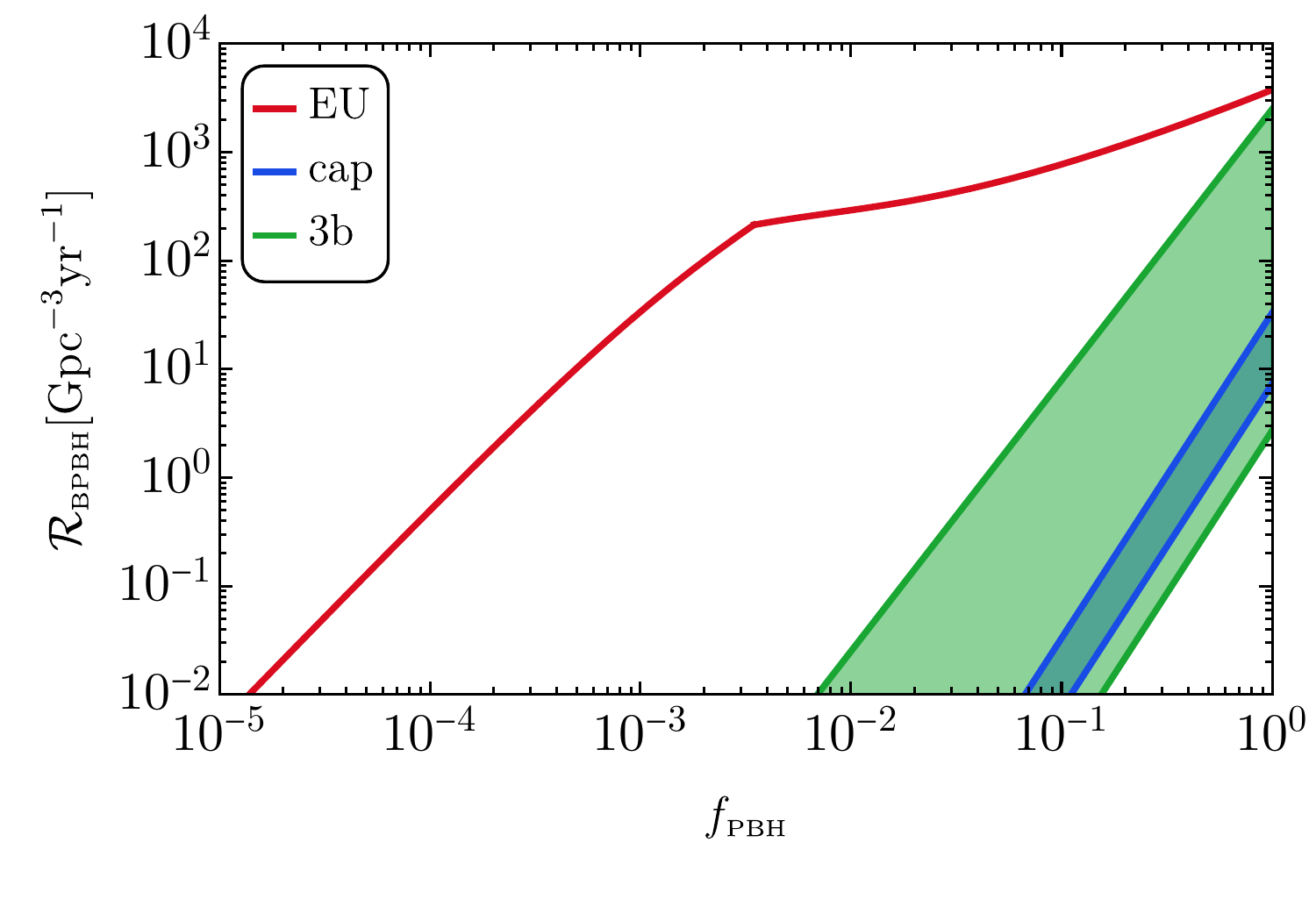}
	\caption{
	Comparison between the merger rate densities of the various channels considered in this work as a function of the abundance $f_\PBH$ for $m = 30 M_\odot$.
	As discussed in the main text, the contribution of 3b binaries is at least comparable to the capture channel, within the uncertainties on $f_c$ and $\gamma$ represented by the colored bands,
	while it can significantly contribute to the overall merger rate only if $f_\PBH$ is of order unity, a value which is ruled out by current constraints in this mass range \cite{Carr:2020gox}. 
	}
	\label{fig:finalcomp}
\end{figure}

The results are summarized in Fig.~\ref{fig:finalcomp} [see also Table~\ref{Tb:parameters} and Eq.~\eqref{eq:diffaccrate}].
As discussed above, the contribution from 3b-induced binaries is comparable to the one from dynamical capture within conservative assumptions, while it becomes significantly larger if one assumes a superthermal distribution of initial eccentricities.
This conclusion is valid independent of the PBH abundance and masses, as indicated by the scaling relation reported in Sec.~\ref{sec:discussion}. 

We compared this scenario with the merger rate of binaries formed in the early Universe. Focusing on the stellar mass range, where the LVKC is currently detecting GW sources, 
we find that the merger rate of 3b binaries cannot significantly contribute to the overall PBH merger rate as PBHs being a dominant fraction of the dark matter is ruled out by current constraints \cite{Carr:2020gox} which force the PBH abundance to be $f_\PBH \lesssim 10^{-3}$ above the solar mass \cite{Ali-Haimoud:2017rtz,Vaskonen:2019jpv,Hall:2020daa,Wong:2020yig}.
We gauged how each channel would contribute when we assume smaller values of $f_\PBH$ or smaller masses (in the asteroid mass range), finding that the early Universe merger rate would be dominant in both cases.
Therefore, given current constraints on the PBH abundance, our results confirm 
the merger rate of binaries formed in the early Universe is the dominant channel 
in the standard PBH formation scenario.

In this study we took a monochromatic PBH mass distribution.
It would be interesting to extend the analysis by considering a broader distribution, even though we expect our results to be marginally affected. We note that such a choice would likely increase the merger rate from the 3b channel, due to the strong dependence of the 3b rate on mass [cf. Eq.~\eqref{eq.3bRate}]. 
Moreover, a mass spectrum in clusters accelerates core collapse due to mass segregation, and thus the more massive PBHs would sink in a smaller region close to the center, boosting interaction rates.

We conclude by reiterating that in the solar mass range large values of the PBH abundance 
are ruled out by bounds from microlensing~\cite{Silk:2022cck,Petac:2022rio,Gorton:2022fyb} 
and GW \cite{Wong:2020yig,Hutsi:2020sol,Franciolini:2021tla} observations.
It was proposed that modified (clustered) initial conditions may help with
evading some of the constraints (see e.g. Ref.~\cite{Atal:2020igj}).
However, this does not represent a viable solution in the LVKC mass range
as PBH clusters would induce large isocurvature perturbations at the scales constrained 
by Lyman-$\alpha$ observations~\cite{DeLuca:2022uvz}. 
Still, such scenarios may be explored assuming smaller values of the abundance and/or lighter mass ranges.   
Within this context, it was suggested that initial clustering would 
enhance the rate of binary disruption in the early Universe~\cite{Raidal:2018bbj,Atal:2020igj,Jedamzik:2020omx}, 
thus reducing their contribution to PBH mergers in the late-time Universe. 
On the contrary, as initial clustering would inevitably boost the rate of binary formation 
from 3b interactions, and further binary-single interactions in clusters cannot suppress 
the merger rate below that estimated adopting the thermal eccentricity distribution, 
we anticipate that this channel should play a key role. 
We leave the study of 3b rates within initially clustered scenarios for future work. 

\acknowledgments
We thank H. Veerm$\ddot{\rm a}$e for useful discussions and comments on the draft.
We also thank V.~Strokov, V.~Baibhav and N.~Stone for discussions.
G.F. thanks Johns Hopkins University for the kind hospitality during the completion of this project.
K.K. and E.B. are supported by NSF Grants No. AST-2006538, No. PHY-2207502, No. PHY-090003, and No. PHY20043 and NASA Grants No. 19-ATP19-0051, No. 20-LPS20-0011, and No. 21-ATP21-0010. 
This research project was conducted using computational resources at the Maryland Advanced Research Computing Center (MARCC).
This work has received funding from the European Union’s Horizon 2020 research and innovation programme under the Marie Skłodowska-Curie Grant Agreement No. 690904.
G.F. acknowledges financial support provided under the European Union's H2020 ERC, Starting Grant Agreement No.~DarkGRA--757480, under the MIUR PRIN, FARE programmes (GW-NEXT, CUP:~B84I20000100001) and 
H2020-MSCA-RISE-2020 GRU.

\appendix

\section{THE ROLE OF BINARY HARDENING}\label{app:hard}

We have discussed the formation of hard binaries in dense PBH clusters through 3b encounters. Such binaries are expected to undergo multiple binary-single interactions before they merge. Then according to the Heggie-Hills law \cite{1975MNRAS.173..729H,1980AJ.....85.1281H}, those binaries tend to become harder in the collisional environment of dense clusters, and their survival probability is extremely close to $100\%$ \cite{1993ApJ...403..271G}.

Hardening is the process by which hard binaries increase their binding energy with time as they interact with a third single compact object in the same environment. As such, their inspiral is accelerated at a constant rate, while their eccentricity grows in a statistical sense \cite{Sesana:2006xw}. Hardening could aid in boosting the $Q$ factor we calculated in Sec.~\ref{sec:fraction}. 
As binaries tighten, at some point their semimajor axis becomes so small and eccentricity increases to such large values that gravitational radiation reaction starts dominating, and takes over the evolution of the binary. 
However, the timescale for a binary to harden enough for GW evolution to dominate is of the order of \cite{Kritos:2020wcl}
\begin{align}
    \tau_{\rm hard}&\simeq335{\rm Gyr}\left({\sigma_v\over{\rm km/s}}{{\rm pc}^{-3}\over n}{15\over H}\right)^{4\over5}\left({m\over30M_\odot}\right)^{-{7\over5}}\nonumber\\&\times{(1-e^2)^{7\over10}}\left(1+{73\over24}e^2+{37\over96}e^4\right)^{-{1\over5}}\,,
\end{align}
which is larger than the Hubble time for $e<0.99$ and becomes $\tau_{\rm hard} \simeq 3.2$~Gyr (still larger than the typical evaporation time for a light cluster) for $e=0.999$.

The symbol $H$ above denotes the hardening rate, a dimensionless number in the range 
$\simeq [15, 20]$. 
Because 3b binary formation is most important in light systems with small escape speeds, the majority of the binaries are expected to be ejected from the minihalo at some point before they harden enough to merge within the cluster \cite{2001CQGra..18.3977L,Antonini:2016gqe}.
Furthermore, the definition of a hard binary depends on the environmental parameters, in this case on the number of objects in the minihalo.
As the cluster's parameters change with time\footnote{For instance, as smaller halos become engulfed into larger ones during the hierarchical assembly, some BPBHs may find themselves in the environment of a larger halo with different velocity dispersion. See Sec.~\ref{sec:halomassfunction} for more details.}, a binary that is initially marginally hard may become soft. However, our assumption of considering $x>5kT$ is a strong one, in the sense that the temperature of the environment does not vary strongly with $N$ (the number of PBHs in the cluster), and the survival probability of the binary does not drop well below unity.
We conclude that the effects of hardening can be safely neglected for the regime of cluster masses in which we are interested.

\bibliography{refs}

\end{document}